\documentclass[conference]{IEEEtran}
\IEEEoverridecommandlockouts

	\usepackage[turkish]{babel}
	\usepackage[utf8]{inputenc} 
	\usepackage[T1]{fontenc}
	\usepackage{caption}

\usepackage{cite}

\ifCLASSINFOpdf
  \usepackage[pdftex]{graphicx}
\else
\fi
\usepackage[cmex10]{amsmath}
\usepackage{multirow}
\usepackage{array}
\usepackage[lofdepth,lotdepth]{subfig}
\usepackage{xfrac}

\AtBeginDocument{%
  \renewcommand\tablename{TABLO}
}

\AtBeginDocument{%
  \renewcommand\abstractname{Abstract}
}

\begin{document}

\IEEEpubid{\makebox[\columnwidth]{978-1-5386-1501-0/18/\$31.00 \copyright\ 2018 IEEE\hfill}
\hspace{\columnsep}\makebox[\columnwidth]{}}
%
\title{Hata Yayılımının En İyi Röle Seçimli İşbirlikli Haberleşme Sistemlerinin Hata Başarımına Etkisi \\Effect of the Error Propagation on the Error Performance of Cooperative Communications with the Best Relay Selection Schemes}
\author{\IEEEauthorblockN{\textit{Ezgi ŞANLI, Ferdi KARA, Hakan KAYA
}}
\IEEEauthorblockA{Elektrik Elektronik Mühendisliği 
\\Bülent Ecevit Üviversitesi\\ Zonguldak, Türkiye \\
ezgisanli93@gmail.com, \{f.kara,hakan.kaya\}@beun.edu.tr
\\
}
}
%
\maketitle
\begin{ozet}
Bu çalışmada en iyi röle seçimli işbirlikli haberleşme sistemlerinin hata başarımları, röleden hedefe doğru hata yayılımı varlığında incelenmiştir. Hata yayılımı ifadesi, ortamda bulunan \textit{M} adet röle arasından en iyi rölenin seçilmesi durumu için literatürde ilk kez türetilmiştir. Elde edilen uçtan uca BHO ifadesi bilgisayar benzetimleri ile desteklenmiştir. Hatasız çözmenin (perfect decoding) aksine hata yayılımı varlığında en iyi rölenin seçilmesinin \textit{M}+1 çeşitlilik derecesini sağlamadığı gösterilmiştir. Bunun yanı sıra, eşik değer seçiminin de sistemin hata başarımı üzerinde önemli bir etkiye sahip olduğu kanıtlanmıştır.
\end{ozet}
\begin{IEEEanahtar}
İşbirlikli Haberleşme, Röle seçimi, Hata yayılımı.
\end{IEEEanahtar}
\begin{abstract}
In this paper, error performance of the cooperative communication systems with the best relay selection scheme is investigated in the presence of error propagation from role to destination. The error propagation expression is derived firstly when the best relay is selected within \textit{M} relays. The derived end-to-end BER expression is verified with the computer simulations. It is shown that the best relay selection does not ensure \textit{M}+1 diversity order under the error propagation unlike perfect decoding. In addition, the threshold selection for the relays has dominant effect on the error performance of the system.
\end{abstract}
%
\IEEEpeerreviewmaketitle
\IEEEpubidadjcol
\section{G{\footnotesize İ}r{\footnotesize İ}ş}
Gelecek nesil Radyo Erişim (Future Radio Access -FRA) ağlarının (5G ve sonrası) çok yüksek hız/kapasite (ver high rate), çok düşük gecikme (ultra-low latency), kitlesel bağlantı (massive connectivity), yüksek gezginlik (very high mobility) ve geniş kapsama (ultra-wide coverage) alanı gibi isterleri karşılaması beklenmektedir [1], [2]. Bu isterlerin karşılanabilmesi adına 5G ve sonrası için önerilen fiziksel seviye teknikler: Kitlesel Çok Girişli-Çok Çıkışlı Sistemler (massive MIMO), Dikgen Olmayan Çoklu Erişim (NOMA), İşbirlikli haberleşme (Cooperative Communcation) ve Milimetre dalga haberleşmesi (Milimeterwave Communication) olarak öne çıkmaktadır [3]–[5]. Bu tekniklerden işbirlikli haberleşme özellikle çok geniş kapsama alanı ve yüksek gezginlik sunma potansiyeline sahiptir. Ayrıca, önerilen diğer teknikler ile birlikte kullanılabilmesi de önemli bir avantaj olarak gözükmektedir. Özellikle, işbirlikli haberleşmede yaşanan kapasite kaybının NOMA kullanılması ile önüne geçilebilmesi nedeniyle, bu iki tekniğin beraber kullanıldığı uygulamalar araştırmacıların ilgisini çekmektedir [6]–[8]. İşbirlikli haberleşme ayrıca akıllı şehir (smart city) uygulamaları için de son derece uygun bir yöntem olarak öne çıkmaktadır [9]. Özellikle araçlar arası iletişim için işbirlikli haberleşmenin kullanılması literatürde birçok çalışmaya konu olmuştur [10]–[12].

İşbirlikli haberleşme fiziksel kısıtlamalar nedeniyle fazla sayıda anten yerleştirilememesi durumunda etkin bir çeşitleme tekniği olarak ortaya çıkmıştır [13]. Bu özelliği sayesinde işbirlikli haberleşme, sensör ağları (sensor networks) gibi tasarsız ağlarda (Ad-Hoc networks) kullanılan bir teknik olmuştur [14].  İşbirlikli ihaberleşme kaynaktan hedefe direkt olarak gönderilen sembole ek olarak aynı sembolün bir röle yardımıyla kaynak-röle ve röle-hedef yolunu izleyerek hedefe bağımsız bir yoldan daha ulaştırılması prensibine dayanmaktadır. Rölede gerçekleştirilen işleme göre işbirlikli haberleşme temel olarak iki grupta incelenmektedir. Bunlar: Kuvvetlendir-Aktar (Aplify- Forward -AF) ve Çöz-Aktar (Decode-Forward -DF) protokolleridir . Bu çalışmanın konusu olan DF protokolü kullanan işbirlikli haberleşmede röle, kaynaktan alınan modüleli sembolü çözüp tekrar modüle ettikten sonra hedefe göndermektedir. DF protokolünün en büyük dezavantajı, kaynak-röle arasındaki sönümlemeden dolayı rölede çözme sırasında hata yapılması ve bu hatalı sembollerin hedefe iletilmesidir. Bu durum literatürde hata yayılımı (error propagation) olarak adlandırılmaktadır [15]. Ortamda birden çok röle bulunması durumunda tüm röleleri kullanmak yerine rölelerden en iyisinin seçilmesi fikri [16]’da önerilmiştir. Böylece tüm rölelerin kullanılması durumu ile aynı çeşitleme derecesi kaynakların daha verimli kullanılmasıyla elde edilebilmekte ve daha yüksek bir hıza/kapasiteye ulaşılabilmektedir. Ayrıca en iyi $N$ adet rölenin seçilmesi durumunda da sistem başarımı araştırılmıştır [17]. Fakat, DF protokolü kullanan röle seçimli sistemler için rölede/rölelerde yapılan hatalar göz ardı edilmiştir. Bu durum sönümleme etkisi göz önünde bulundurulduğunda çok makul bir yaklaşım değildir. 

Bu çalışmada DF protokolü kullanan $M$ adet röleden eşik-değer tabanlı olarak en iyi rölenin seçildiği durumda sistemin hata başarımı, hata yayılımının varlığında incelenmiştir. Röleden hedefe doğru oluşan hata yayılımı altında hedefteki Bit Hata Oranının (BHO) ifadesi analitik olarak elde edilmiştir.  Hata yayılımının sistemin çeşitlilik derecesine etkisi incelenmiştir. Ayrıca eşik-değer tabanlı röle seçim kuralı için eşik değerin ne olması gerektiği tartışılmış ve optimum eşik değer için nümerik sonuçlar sunulmuştur. II. Bölümünde sistem modeli ve röle seçim kuralı tanıtılmıştır. III. Bölümde sistemin uçtan uca BHO ifadesi hata yayılımı varlığında türetilmiştir. Daha sonra, elde edilen analitik ifadeler bilgisayar benzetimleri ile desteklenerek, hata yayılımının sistem başarımına ve çeşitlilik derecesine etkisi farklı senaryolar için IV. Bölümde sunulmuştur. Son olarak, V. Bölümde sonuçlar tartışılarak çalışma sonlandırılmıştır.
\section{S{\footnotesize İ}stem Model{\footnotesize İ}}
Bu çalışmada ortamda bulunan M adet röle arasından seçilen en iyi link kalitesine sahip röle yardımıyla kaynak ile hedef arasındaki iletişimin iki atlamalı olarak sağlandığı işbirlikli iletişim sistemi ele alınmıştır. Her birim, tek antenli yarı çift yönlü (half-duplex) yapıya sahiptir ve toplam iletişim iki zaman diliminde tamamlanmaktadır. Hedefte iki zaman dilimi sonunda alınan işaretler Maksimum Oranlı Birleştirme (Maximum Ratio Combining -MRC) kullanılarak birleştirilmektedir. Herhangi iki birim arasındaki kanal sönümleme katsayısının zarfı Rayleigh rasgele değişkeni olarak modellenmiştir. Sistemdeki tüm röleler DF protokolünü kullanmaktadır. Modülasyon işaret yıldız kümesi olarak İkili Faz Kaydırmalı Anahtarlama (Binary Phase Shift Keying -BPSK) kullanılmıştır. Röle seçim kuralı olarak eşik değer tabanlı (threshold-based) röle seçim kuralı kullanılmıştır. Bu sistemde röle seçimi iki aşamada tamamlanmaktadır. İlk aşamada kaynaktan $i.$ röleye gelen sembolün İşaret Gürültü Oranı (İGO) değeri olan $\gamma_{{sr}_i}$’ye bakılmaktadır. Kaynaktan röleye gelen sembolün İGO değeri, belirli bir eşik değerinden  ($\gamma_{th}$) küçük ise rölenin sembolü doğru çözemediğine karar verilir ve bu röle seçim kümesi olarak adlandırılan $C$ kümesinin dışında bırakılır. İGO değerinin eşik değerinden büyük olduğu rölelerin ($\gamma_{sr_i}>\gamma_{th}$) ise gelen sembolü doğru çözdüğü varsayılır ve bu röle $C$ kümesinin içerisinde yer alır. İkinci aşamada ise $C$ kümesi içerisinden röle-hedef arasındaki İGO değeri en yüksek olan röle ($max{\left\{\gamma_{rd_i}\right\}},\ i\in C$) seçilir. En iyi röle belirlendikten sonra seçim dışında kalan röleler diğer sembole kadar bekleme konumuna geçmektedir. 

İlk zaman diliminde hedef ve $i$. röle tarafından alınan işaretler [13]
\begin{equation}
\begin{split}
&\ y_{sd}=\sqrt{P_s}\ h_{sd}\ x+n_{sd}\\
&\ y_{sr_i}=\sqrt{P_s}\ h_{{sr}_i}\ x+n_{{sr}_i}
\end{split}
\end{equation}
olur. Burada; $x$ vericiden gönderilen modülasyonlu sembolü, $P_s$ verici gücünü, $n_{sd}$ ve $n_{{sr}_i}$ sırasıyla kaynak-hedef ve $i$. röle-hedef arasındaki çift yönlü güç spektral yoğunluğu $N_0/2$ olan Toplanır Beyaz Gauss Gürültüsünü, $h_{sd}$ ve $h_{{sr}_i}$ sırasıyla kaynak-hedef ve kaynak- $i$. röle arasındaki ${\sigma^2}_{sd}$ ve  ${\sigma^2}_{{sr}_i}$  varyanslı sönümleme katsayılarını ifade etmektedir. İkinci zaman diliminde ise seçilen röleden hedefe gönderilen işaret
\begin{equation}
\begin{split}
y_{rd}=\sqrt{P_r}\ h_{r^1d}\ x_r+n_{rd}
\end{split}
\end{equation}
$x_r$,  DF protokolü kullanan rölenin kaynaktan gelen sinyali çözüp tekrar modülasyon işlemine tabi tutması sonucunda oluşan sembolü, $h_{r^1d}$  seçilen röle-hedef arasındaki ${\sigma^2}_{r^1d}$  varyanslı sönümleme katsayısını, $n_{r^1d}$ çift yönlü güç spektral yoğunluğu $N_0/2$ olan toplanır beyaz Gauss gürültüsünü ifade etmektedir.
Kanalın hedefte bilindiği varsayımı altında MRC yöntemi kullanılarak çıkış İGO değeri maksimize edilir ve hedefte alınan işaret [18]
\begin{equation}
\begin{split}
y=a_1\ y_{sd}+a_2\ y_{rd}
\end{split}
\end{equation}
Burada $a_1$ ve $a_2$ MRC katsayılarıdır, $a_1=\frac{\sqrt{P_s\ }{h^\ast}_{sd}}{N_0}$ ve $a_2=\frac{\sqrt{P_{r}\ }{h^\ast}_{r^1d}}{N_0}$  olarak tanımlanır.
\section{S{\footnotesize İ}stem{\footnotesize İ}n Uçtan Uca Hata {\footnotesize İ}fades{\footnotesize İ}n{\footnotesize İ}n Eldes{\footnotesize İ}}
Eşik değer tabanlı yöntem ile seçilen en iyi link kalitesine sahip röle üzerinden yapılan iletişimde hedefte meydana gelen uçtan uca hata olasılığı ifadesi
\begin{equation}
\begin{split}
P_{e2e}=\sum_{i=0}^{M}{P_r\left(N_r=i\right)P_r(\varepsilon_{e2e}|N_r=i)}
\end{split}
\end{equation}
olarak verilir [19]. Burada $P_r(\varepsilon_{e2e}|N_r=i)$, $M$ adet rölenin bulunduğu ortamda $C$ kümesine dahil olan röle sayısının $i$ adet olma olasılığı altında oluşan hata olasılığını, $P_r\left(N_r=i\right)$ ise $M$ adet rölenin bulunduğu ortamda $C$ kümesinde yer alan röle sayısının $i$ adet olma olasılığını göstermektedir ve bu olasılık 
\begin{equation}
\begin{split}
&P_r\left(N_r=i\right)=\binom{M}{i}{P_{dec}}^i{(1-P_{dec})}^{M-i}, i=0,1,...M
\end{split}
\end{equation}
şeklinde tanımlanır [19]. $P_{dec}$ röleye gelen sembolün İGO değerinin eşik değeri $\gamma_{th}$’den büyük olma olasılığını belirtmektedir. 
\begin{equation}
\begin{split}
P_{dec}=Pr\left({\overline{\gamma}}_{sr}>\gamma_{th}\ \right)=e^{-\sfrac{\gamma_{th}}{{\overline{\gamma}}_{sr}}}
\end{split}
\end{equation}
Burada  ${\overline{\gamma}}_{sr}$, kaynak - röle arasındaki ortalama İGO’dur ve  $ {\overline{\gamma}}_{sr}=E\left[\gamma_{sr}\right]$ olarak verilmektedir. $E\left[.\right]$ ortalama operatörüdür. $\gamma_{sr}$ ise kaynak - röle arasındaki anlık İGO’dur ve $\gamma_{sr}=\sfrac{P_s\left|h_{sr}\right|^2}{N_0}$ olarak verilmektedir. 
Denlem (5) Denklem (4)'de yerine konulursa
\begin{equation}
\begin{split}
P_{e2e}=&{(1-P_{dec})}^M P_{non-coop}\\
&+\sum_{N_r=1}^{M}\binom{M}{N_r}{P_{dec}}^{N_r}\left(1-P_{dec}\right)^{M-N_r}P_{div}             
\end{split}
\end{equation}
elde edilir. Burada $N_r$, $C$ kümesine dahil olan röle sayısıdır
$P_{non-coop}$ $C $ kümesine hiçbir rölenin girmediği yani $N_r=0$ durumundaki ortalama hata olasılığını ifade etmektedir. $P_{div}$ ifadesi $C$ kümesi içerisinden en iyi link kalitesine sahip röle seçildiği durumdaki işbirlikli iletişimin ortalama hata olasılığını ifade etmektedir ve
\begin{equation}
\begin{split}
P_{div}=P_{SR}\times P_{prop}+\left(1-\ P_{SR}\right)\times P_{coop}            
\end{split}
\end{equation}
olarak tanımlanır [20]. $C$ kümesine hiçbir rölenin girmediği durumundaki ortalama hata olasılığını ifadesi
\begin{equation}
\begin{split}
P_{non-coop}=\frac{1}{\pi}\int_{0}^{\sfrac{\pi}{2}}{M_{\gamma_{sd}}\left(\sfrac{1}{{sin}^2(\theta)}\right)\ d\theta} 
\end{split}
\end{equation}
olarak verilir [21]. Burada $M(.)$ Moment üreten fonksiyon olarak tanımlanır Rayleigh sönümlemeli kanalda oluşan İGO ifadesinin MGF’si $M_{\gamma_{sd}}\left(s\right)=\left(1+\overline{\gamma}_{sd}s\right)^{-1}$ olarak verilir [21]. $P_{SR}$ , rölenin kaynaktan gelen sembolü yanlış çözmesinin olasılığı; bir başka ifade ile eşik değerinin kötü seçimine bağlı meydana gelen hata yayılımının oluşma olasılığıdır. Bu hata yayılımı sonucunda MRC sonrası hedefte meydana gelen hata olasılığı $P_{prop}$ olarak adlandırılmaktadır. [20]’de yazarlar Rayleigh sönümlemeli kanal için
\begin{equation}
\begin{split}
&P_{SR}\left(\gamma_{th},{\overline{\gamma}}_{sr}\right)\le\\ 
&Q\left(\sqrt{2\ \gamma_{th}}\right)-e^\frac{\gamma_{th}}{{\overline{\gamma}}_{sr}}\sqrt{\frac{1}{1+\frac{1}{{\overline{\gamma}}_{sr}}}Q\left(\sqrt{2\gamma_{th}\left(1+\sfrac{1}{{\overline{\gamma}}_{sr}}\right)}\right)}  
\end{split}
\end{equation}
olarak vermişlerdir. $P_{coop}$ hatasının hesabı için, kaynak-hedef arasındaki MGF ifadesi ve kaynak-röle arasındaki MGF ifadesi yazılmalıdır. MGF ifadelerini kullanarak $P_{coop}$ ifadesi 
\begin{equation}
\begin{split}
P_{coop}=\frac{1}{\pi}\int\limits_{0}^{\sfrac{\pi}{2}}{M_{\gamma_{sd}}\left(\frac{1}{{sin}^2(\theta)}\right)M_{r^1d}\left(\frac{1}{{sin}^2(\theta)}\right)d\theta}
\end{split}
\end{equation}
olarak yazılır. $M_{r^1d}$, $N_r$ adet röleden en yüksek İGO değerine sahip rölenin İGO’sunun MGF’si olarak tanımlanır [22] ve
\begin{equation}
\begin{split}
M_{\gamma_{r^1d}}\left(s\right)=N_r\sum_{k=0}^{N_r-1}{{(-1)}^k\binom{N_r-1}{k}\frac{1}{k+1+{\overline{\gamma}}_{r^1d}\ s}}
\end{split}
\end{equation}
şeklinde verilir. 

Hata yayılımı sonucunda alıcıdaki hatayı bulmak adına, BPSK için rölenin hatalı çözme yaptığı varsayımı altında, vericiden gönderilen sembol $x=1$ ise röleden hedefe gönderilen sembol $x=-1$ olacaktır. Kaynaktan gönderilen (direkt yol) ve röleden gelen sembollerin MRC ile birleştirilmesi sonucunda alıcıda alınan işaret
\begin{equation}
\begin{split}
y=&\left(\frac{\left|h_{sd}\right|^2\ P_{sd}}{N_0}-\frac{\left|h_{r^1d}\right|^2P_{rd}}{N_0}\right)\\
&+\frac{{h^\ast}_{sd}\sqrt{P_{sd}}}{N_0}\ n_{sd}+\frac{{h^\ast}_{r^1d}\ \sqrt{P_{rd}}}{N_0}\ n_{rd}\\ 
 =&\left(\gamma_{sd}-\gamma_{r^1d}\right)+\widetilde{n}            
\end{split}
\end{equation}
olarak bulunur [23]. Burada $\widetilde{n}$ sıfır ortalamalı ve  $\frac{1}{2}\left(\gamma_{sd}+\gamma_{r^1d}\right)$ varyanslı Gauss rasgele değişkenidir. Vericiden $x=1$ gönderilmesi durumunda alıcıda hatanın oluşabilmesi için karar kuralı $y<0$ olmalıdır. $\gamma_{sd}$ ve $\gamma_{rd}$ anlık değerler olduğu için ortalama İGO’lara göre, $P_{prop}$ olasılığının ortalaması 
\begin{equation}
\begin{split}
\overline{P}_{prop}=&\int_{0}^{\infty}\int_{0}^{\infty}{Q\left(\frac{\gamma_{sd}-\gamma_{r^1d}}{\sqrt{\sfrac{\left(\gamma_{sd}+\gamma_{r^1d}\right)}{2}}}\right)}\\
&\times{f_{\gamma_{sd}}\left(\gamma_{sd}\right)f_{\gamma_{r^1d}}(\gamma_{r^1d}) d\gamma_{sd} d\gamma_{r^1d}}          
\end{split}
\end{equation}
şeklinde verilir. Denklem (14)’de verilen ifadenin kapalı formda elde edilmesi zor olduğundan, Denklem (13)'e göre $\gamma_{sd}-\gamma_{r^1d}<0$ olması durumunda hata meydana geldiği yaklaşımı ile hata yayılımı $P_{prop}\cong\ P{(\gamma}_{sd}-\gamma_{r^1d}<0\ |{\ \overline{\gamma}}_{sd},\ {\overline{\gamma}}_{r^1d})$ olarak verilebilir [23]. Bu durumda 
\begin{equation}
\begin{split}
P_{prop}\cong\int_{0}^{\infty}\int_{0}^{\gamma_{r^1d}}{f_{\gamma_{sd}}\left(\gamma_{sd}\right)f_{\gamma_{r^1d}}(\gamma_{r^1d}) d\gamma_{sd} d\gamma_{r^1d}}          
\end{split}
\end{equation}
olarak verilebilir. $f_{\gamma_{r^1d}}(\gamma_{r^1d})$ ifadesi $C$ kümesine giren $N_r$ adet röleden en yüksek İGO değerine sahip rölenin İGO’sunun olasılık yoğunluk fonksiyonu olarak ifade edilir ve [24]'te verilen sıra istatistiğinden yararlanılarak
\begin{equation}
\begin{split}
f_{\gamma_{r^1d}}\left(\gamma_{r^1d}\right)=N_r\left[F_{\gamma_{rd}}(\gamma_{rd})\right]^{N_r-1}f_{\gamma_{rd}}\left(\gamma_{rd}\right)         
\end{split}
\end{equation}
şeklinde tanımlanır [23]. Burada $f_{\gamma_{rd}}\left(\gamma_{rd}\right)$  ve $F_{\gamma_{rd}}(\gamma_{rd})$ sırasıyla, $M$ adet sıralanmamış özdeş bağımsız rasgele değişkenlerin olasılık yoğunluk ve olasılık dağılım fonksiyonlarını ifade etmektedir. Denklem (16)'da verilen ifade ve kaynak hedef arasındaki Rayleigh sönümlemeli kanal için İGO’nun olasılık yoğunluk fonksiyonu olan üstel dağılım Denklem (15)’te yerine konulur ise
\begin{equation}
\begin{split}
P_{prop}=&\sum_{k=0}^{M-N_r}{M\ \binom{M-1}{N_r-1} \left(-1\right)^k}\binom{M-N_r}{k}\\
&\times\left(\frac{-{\overline{\gamma}}_{sd}}{{\overline{\gamma}}_{r^1d}+(k+1){\overline{\gamma}}_{sd}}+\frac{1}{k+1}\right)    
\end{split}
\end{equation}
olarak bulunur. Sistemin uçtan uca BHO başarımı Denkelem (6),(9),(10),(11),(17) nin Denklem (7)'de yerine konulması ile elde edilir.
\section{Nümer{\footnotesize İ}k Sonuçlar}
Şekil 1’de hata yayılımının etkisi farklı eşik değerler seçilmesi durumunda irdelenmiştir. Ortamda $M=4$ adet röle bulunması durumunda eşik değer tabanlı olarak en iyi röle seçilmesi durumunda sistemin hata başarımı eşik değer $\gamma_{th}=0dB, 5dB$, $8dB$ ve $\gamma_{opt}$ için sunulmuştur. $\gamma_{opt}$ değerleri Denklem (7)'de verilen uçtan uca BHO ifadesini minimize eden değerlerin nümerik olarak hesaplanması ile elde edilmiştir. $\gamma_{opt}$ değerleri $M=4$ için Tablo I'de verilmiştir. Rölenin verici gücü ile kaynak verici gücünün eşit olduğu varsayılmıştır $\left(P_r=P_s \right)$. Tüm röleler için $h_{{sr}_i}$'lerin ve $h_{{rd}_i}$'lerin kendi aralarında özdeş ve bağımsız Rayleigh zarfına sahip rastgele değişkenler olduğu varsayılmıştır. Ortalama kanal güçleri  $\sigma^2_{sd}=-3dB$, $\sigma^2_{{sr}_i}=0dB$\ ve\ $\sigma^2_{r^1d} =0dB$ olarak alınmıştır. Sistemin hata başarımı toplam İGO ($\sfrac{\left(P_r+P_s\right)}{N_0}$)’ya göre verilmiştir. 
\begin{figure}[htbp]
	\centering
	\shorthandoff{=}  
	\includegraphics[height=4.9cm, width=7cm]{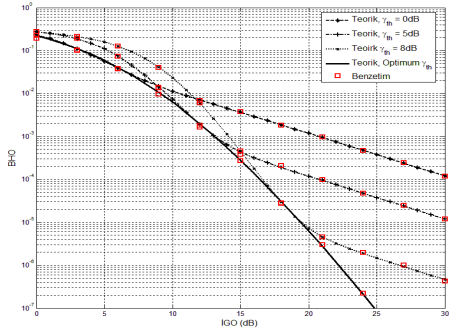}
	\shorthandon{=} 
	\caption{\footnotesize Farklı Eşik Değerleri için Hata Yayılımı Altında BHO Başarımı}
	\label{sekil1}
\end{figure}

Şekil 2’de sistem çeşitlilik derecesini incelemek adına $M=3,4,5$ olması durumunda sistem hata başarımı incelenmiştir. $\gamma_{th}=5dB$ olarak alınmış diğer tüm şartlar Şekil 1’de verilen sanaryoda olduğu gibi kabul edilmiştir. Hata yayılımının sistem çeşitlilik derecesine etkisinin anlaşılabilmesi için $C$ kümesine seçilen rölelerde hata yapılmaması durumu (hatasız çözme -perfect decoding) için benzetimler de sunulmuştur.
\begin{table}[ph]
  \centering
  \caption{\textsc{$M=4$ Röle {\footnotesize İ}ç{\footnotesize İ}n Opt{\footnotesize İ}mum Eş{\footnotesize İ}k Değerler}}
  \label{tablo1}
  \begin{tabular}{|c|c|c|c|c|c|c|c|c|c|}
    \hline
    \multirow{2}{*}{} & \multicolumn{9}{c|}{Toplam İGO(dB)} \\
    \cline{2-10}
    &0&3&6&9&12&15&18&21&24 \\
    \hline
     $\gamma_{opt}$(dB) &-7,9&-3,9&-0,7&1,9&4,0&5,7&7,25&8,4&9,5\\
    \hline
  \end{tabular}
\end{table}
\begin{figure}[htbp]
	\centering
	\shorthandoff{=}  
	\includegraphics[height=4.9cm, width=7cm]{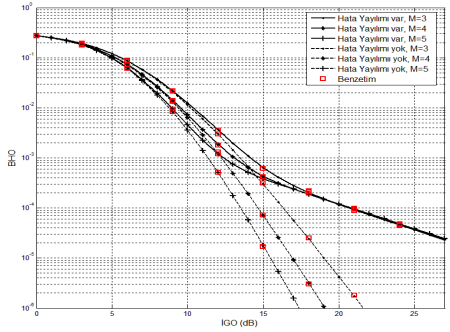}
	\shorthandon{=} 
	\caption{\footnotesize Hata Yayılımı Varlığında ve Yokluğunda Sistem BHO Başarımı.}
	\label{sekil1}
\end{figure}
\section{Sonuçlar ve Tartışma}
Hata yayılımı, DF protokolü kullanan işbirlikli haberleşme sistemlerinin hata performanslarında önemli rol oynamaktadır. Bu çalışmada hata yayılımı altında eşik değer tabanlı röle seçim kuralı kullanılan işbirlikli haberleşme sisteminin hata başarımı ifadeleri elde edilerek sonuçlar bilgisayar benzetimleri ile karşılaştırılmıştır. Elde edilen analitik ifadeler bilgisayar benzetimleri ile mükemmel uyum göstermektedir. 
Literatürde yapılan çalışmalarda en iyi röle seçimi senaryolarında ortamda $M$ adet röle bulunması durumunda sistemin çeşitlilik derecesinin $M+1$ olacağı söylense de bu durum seçilen rölede çözme sırasında hata yapılmaması şartına (hatasız çözme) bağlıdır. Fakat, kablosuz haberleşme sistemleri için bu yaklaşım kabul edilebilir olmaktan biraz uzaktır. Bu çalışmada hata yayılımı olması durumunda ortamda bulunan $M$ adet röleden en iyisi seçilse bile sistemin çeşitlilik derecesinin $M+1$ olmayacağı gösterilmiştir. Hatta yanlış bir eşik değer seçiminde $(\gamma_{th}\rightarrow-\infty \ dB)$ çeşitlilik derecesi $1$ olacaktır. Bu nedenle eşik değerinin seçimi son derece önem arz etmektedir. Ayrıca benzetim sonuçlarından da görüleceği gibi, farklı İGO bölgelerinde farklı eşik değerler için daha iyi bir hata başarımı elde edilmektedir. Bu durum, optimum eşik değerin kanal güçlerinin yanı sıra İGO değerine de bağlı olarak dinamik bir şekilde seçilmesi gerektiğini göstermektedir. 
\end{document}